# Strongly-bound excitons and trions in anisotropic 2D semiconductors

Sangho Yoon[1,2,†], Taeho Kim[1,2,†], Seung-Young Seo[1,2], Seung-Hyun Shin[3], Su-Beom Song[1,2], B. J. Kim[2,3], Kenji Watanabe[4], Takashi Taniguchi[5], Gil-Ho Lee[3], Moon-Ho Jo[1,2], Diana Y. Qiu[6*], Jonghwan Kim[1,2,3*]

[1] Department of Materials Science and Engineering, Pohang University of Science and Technology, Pohang, Republic of Korea
[2] Center for Artificial Low Dimensional Electronic Systems, Institute for Basic Science (IBS), Pohang, Republic of Korea
[3] Department of Physics, Pohang University of Science and Technology, Pohang, Republic of Korea
[4] Research Center for Functional Materials, National Institute for Materials Science, Tsukuba, Ibaraki, Japan
[5] International Center for Materials Nanoarchitectonics, National Institute for Materials Science, Tsukuba, Ibaraki, Japan
[6] Department of Mechanical Engineering and Materials Science, Yale University, New Haven, CT, USA

[†] These authors contribute equally to this work.
[*] To whom correspondence should be addressed.
[*] Email : jonghwankim@postech.ac.kr ; diana.qiu@yale.edu

**Abstract:**

Monolayer and few-layer phosphorene are anisotropic quasi-two-dimensional (quasi-2D) van der Waals (vdW) semiconductors with a linear-dichroic light-matter interaction and a widely-tunable direct-band gap in the infrared frequency range. Despite recent theoretical predictions of strongly-bound excitons with unique properties, it remains experimentally challenging to probe the excitonic quasiparticles due to the severe oxidation during device fabrication. In this study, we report observation of strongly-bound excitons and trions with highly-anisotropic optical properties in intrinsic bilayer phosphorene, which are protected from oxidation by encapsulation with hexagonal boron nitride (hBN), in a field-effect transistor (FET) geometry. Reflection contrast and photoluminescence spectroscopy clearly reveal the linear-dichroic optical spectra from anisotropic excitons and trions in the hBN-encapsulated bilayer phosphorene. The optical resonances from the exciton Rydberg series indicate that the neutral exciton binding energy is over 100 meV even with the dielectric screening from hBN. The electrostatic injection of free holes enables an additional optical resonance from a positive trion (charged exciton) ~ 30 meV below the optical bandgap of the charge-neutral system. Our work shows exciting possibilities for monolayer and few-layer phosphorene as a platform to explore many-body physics and novel photonics and optoelectronics based on strongly-bound excitons with two-fold anisotropy.

**Main Text:**

**Introduction**

Two-dimensional vdW semiconductors have emerged as a fascinating class of materials for exploring novel excitonic phenomena(*1-4*) and optoelectronic applications(*5*). Such opportunities arise from the combination of spatial confinement and the dramatically enhanced Coulomb interaction due to reduced dielectric screening in the atomically thin structure, which greatly enhances the exciton binding energy compared to bulk materials, as shown for transition metal dichalcogenides (TMD) monolayers(*6*). The exciton binding energy can be on the order of several 100 meV, which is more than an order of magnitude greater than in typical inorganic semiconductors. Higher-order excitonic complexes such as a charged excitons (or trions)(*7, 8*), biexcitons(*9, 10*), and even charged biexcitons(*11-13*) are observed to be stable at elevated temperatures. Furthermore, the strong Coulomb interaction can be readily controlled via the modification of the local dielectric environment, which provides a new pathway to engineer the electrical and optical properties of semiconductors based on many-body interactions(*14-16*).

Monolayer and few-layer phosphorene are unique quasi-2D vdW semiconductors with distinct physical properties. The puckered intralayer structure (Fig. 1A) exhibits a highly anisotropic charge transport and a linear-dichroic optical response with respect to the crystalline axes(*17-20*). As the thickness varies from the monolayer to the bulk limit, the electronic structure maintains a direct-bandgap, while the optical bandgap substantially decreases from ~ 1.7 eV to ~ 0.3 eV due to the strong interlayer interaction(*21, 22*). Such unusual properties of few-layer phosphorene have enabled exciting new possibilities for device applications including widely-tunable infrared optoelectronics(*23, 24*) and extremely low-power transistors(*25*).

However, experimental studies of anisotropic excitonic quasiparticles in monolayer and few-layer phosphorene have been significantly hampered due to the low crystal quality in the limit of atomic thickness(*21, 22, 26-29*). A recent study shows that the crystal can be oxidized by the surface chemical environment of an oxide substrate even in a glovebox with a globally low level of oxygen and moisture(*30*). The degradation from the substrate surface can be suppressed by utilizing a hydrophobic substrate such as polydimethylsiloxane (PDMS)(*28*). Nevertheless, the polymer substrate is limited for device fabrication and low-temperature measurements, which are essential to investigate the structure and properties of excitons and to demonstrate optoelectronic applications. On the other hand, hBN, a vdW layered wide-bandgap insulator (~6 eV), has been widely utilized for two-dimensional materials as an excellent encapsulating material due to its atomically flat and dangling-bond-free surface. The disorder from the external environment including the substrate surface roughness, charge traps, and chemical degradation can be dramatically suppressed by encapsulation with hBN, which has provided experimental access to novel intrinsic physical properties of graphene, TMD monolayers, and air-sensitive 2D materials(*25, 31-33*).

In this study, we report observation of strongly-bound excitons and trions (charged excitons) with highly anisotropic optical properties of bilayer phosphorene in a field-effect transistor (FET) geometry with hBN encapsulation. Reflection contrast and photoluminescence spectroscopy reveal a linear-dichroic optical spectra from a series of anisotropic excitons and trions with remarkably narrow linewidth which allows us to probe the detailed excitonic structure. The optical resonances from the exciton Rydberg series shows that the neutral exciton binding energy in the encapsulated bilayer phosphorene is greater than 100 meV. The electrostatic injection of free holes enables the observation of an additional optical resonance from the positive trion ~ 30 meV below the lowest-energy singlet exciton state. In combination with ab initio calculations of the energy splitting between the spin-singlet and spin-triplet exciton states, we estimate that the binding energy of the positive trions is ~ 20 meV. Despite the large dielectric screening from hBN, bilayer phosphorene shows tightly-bound excitons and trions with remarkably large binding energies.

**Results**

Fig. 1A shows the crystal structure of bilayer phosphorene where phosphorus atoms form a puckered honeycomb lattice with two distinct crystalline axes along the X (armchair) and Y (zig-zag) directions. An optical microscope image and a schematic of the representative FET device are shown in Fig. 1B and 1C, respectively. Two hBN crystals (~ 20 nm thick) encapsulate the bilayer phosphorene to prevent oxidation and avoid the oxide substrate-induced defects including dielectric disorder(*33*). Two flakes of few-layer graphene are in contact with the bilayer phosphorene as a source and a drain electrode. Additional few-layer graphene is placed under the bottom hBN which controls the carrier concentration electrostatically (see Materials and Methods for the detailed fabrication procedures). The source-drain current as a function of the gate voltage ($V_B$) shows the characteristic behavior of a bipolar transistor which indicates that both electrons and holes can be injected in the channel of bilayer phosphorene (see Supplementary Materials). We find that bilayer phosphorene is slightly hole-doped at $V_B = 0$ V presumably due to defects in the crystal as reported in the previous literature(*18, 34*).

We measure the absorption and photoluminescence spectra of the bilayer phosphorene for the charge-neutral case ($V_B$ = 0.2 V). Fig. 2A shows the polarization-resolved reflection contrast spectra ($\Delta R/R$). Linearly-polarized broadband white light from a tungsten-halogen lamp illuminates the sample in a home-built confocal microscope set-up. All our optical measurements are carried out at a temperature of 10 K unless otherwise specified. For atomically thin vdW crystals on a transparent sapphire substrate, $\Delta R/R$ is approximately proportional to the real part of the optical conductivity (i.e. optical absorption)(*21, 35*). A prominent resonance is observed at 1.129 eV for the polarization along the arm-chair direction (Red solid line labelled as X-pol.) of the bilayer phosphorene crystal. On the other hand, $\Delta R/R$ shows an overall broad background for the polarization along the zig-zag direction (Black solid line labelled as Y-pol.). The observed linear-dichroic resonance with a Lorentzian lineshape originates from the excitonic transition at the optical band edge of the bilayer phosphorene as reported in the previous literature(*16, 21, 22, 28, 36, 37*). In particular, a mirror symmetry in the xz-plane of bilayer phosphorene (Fig. 1A) leads to the strictly-forbidden dipole interaction of Y-polarized light at the band edge. The broad background in the spectra for both polarization configurations arises due to the dielectric capping layers of hBN which start introducing the imaginary part of the optical conductivity in $\Delta R/R$(*21, 35*).

Fig. 2B shows the photoluminescence spectra with the unpolarized laser excitation at 1.96 eV. The polarization of the luminescence is analyzed by a combination of a half-wave plate and a linear polarizer in front of the spectrometer. The spectra show perfectly polarized photoluminescence along the armchair direction (Red solid line labelled as X-pol.) as expected from the anisotropic dipole interaction of bilayer phosphorene. In addition, a strong signal is observed at 1.129 eV which is exactly the same energy of the resonance in $\Delta R/R$ as marked with a black dashed line in Fig. 2A and 2B. The excellent agreement between the absorption and the luminescence edges confirms the characteristic of the direct bandgap semiconductor for bilayer phosphorene. The full-width-half-maximum (FWHM) of the luminescence peak is ~ 10 meV. The broad and small signal at lower energy is presumably from defect-related states. The absence of a Stokes shift and the remarkably narrow linewidth indicate that high crystalline quality of bilayer phosphorene can be achieved via hBN encapsulation as demonstrated for graphene and TMD monolayers(*32, 33*).

The high quality of the sample enables detailed investigation of the excitonic transitions at the optical band edge. $\Delta R/R$ (Red solid line in Fig. 2A) shows another pronounced resonance at 1.229 eV. The strong polarization dependence indicates that the resonance originates from the bilayer phosphorene. The optical transition between the higher-lying sub-bands, on the other hand, is located at ~ 2.4 eV which is far away in energy(*16, 21*). Therefore, we attribute the resonances in series at 1.129 eV and 1.229 eV as the optical transitions for the exciton 1s and 2s states, respectively. The assignment is further supported by our GW plus Bethe-Salpeter equation (GW-BSE) calculations(*38-40*), which find an energy difference of 160 meV between the exciton 1s and 2s states in freestanding bilayer phosphorene and 90 meV when the screening from the hBN encapsulation is included with a Rytova-Keldysh type potential(*41, 42*) (see Materials and Methods). We note that the recent study reports similar excitonic transitions in bilayer phosphorene on PDMS at room temperature(*28*).

Due to the large transfer of oscillator strength to the exciton states, the continuum onset is not visible in the reflection contrast spectrum. However, the binding energy of the exciton can be estimated from the excitonic absorption lines in Fig. 2A. In the simplest approach, the isotropic 2D hydrogenic model predicts that the exciton binding energy, $E_B^{X_0}$, should be $E_B^{X_0} = \frac{9}{8}\Delta_{12}$, where $\Delta_{12}$ is the energy difference between the exciton 1s and 2s states. For unencapsulated quasi-2D materials, it is well-established in both theory and experiment that exciton excitation series deviate strongly from the isotropic 2D hydrogenic model due to the inhomogeneous dielectric environment, and the ratio of $E_B^{X_0}$ and $\Delta_{12}$ can be greater than 2 in freestanding quasi-2D materials(*16, 43-46*). However, in the presence of encapsulation, if the dielectric constant of the encapsulating material is similar to that of the functional material, as is the case for hBN and few-layer phosphorene(*16, 47, 48*), the dielectric environment becomes more homogeneous and is better approximated by the 2D hydrogenic model. Based on our experimentally measured $\Delta_{12}$ of ~100 meV, we use the 2D hydrogenic model to estimate a binding energy of ~110 meV. A more accurate theoretical model (see Materials and Methods), which was fit to our GW-BSE calculation and includes the anisotropy of the band structure, gives a similar value of 98 meV for the exciton binding energy of bilayer phosphorene encapsulated in hBN.

We investigate the effect of doping on the exciton structure by applying a gate voltage which electrostatically controls the carrier concentration in bilayer phosphorene. Fig. 3A and 3D show photoluminescence spectra and ΔR/R with polarization along the armchair direction, respectively, for the hole-doped case. In both the luminescence and absorption spectra, the exciton 1s peak at 1.129 eV diminishes, and the new peak emerges at ~ 1.1 eV as the gate voltage is varied from 0.2 V (the charge-neutral case) to -1.5 V (the hole-doped case). The emerged optical resonance also shows strong polarization dependence. At $V_B$ = -1.5 V, as a representative example for the hole-doped case, the magnitude of the peak at ~1.1 eV exhibits a $cos^2\theta$ pattern (blue empty circles in Fig. 3C and blue filled circles in Fig. 3F for photoluminescence and ΔR/R, respectively) in the function of the light polarization angle $\theta$. The negligible optical response at $\theta = 90°$ (polarization along the zig-zag direction) indicates that the dipole interaction for Y-polarization is forbidden, identical to the optical resonance at ~1.129 eV for the charge neutral system (Grey-empty circles in Fig. 3C and Grey-filled circles in Fig. 3F). The large oscillator strength of the new peak in ΔR/R implies that the new peak does not originate from long-lived localized defect states. Instead, our optical spectra suggest that a new delocalized state with strong anisotropy emerges due to the presence of free holes. The gate-dependent photoluminescence spectra are also reported in previous studies of monolayer and trilayer phosphorene on a $SiO_2$/Si substrate without encapsulation(*27, 29*).

The emerged resonance in the photoluminescence and ΔR/R spectra can be nicely explained by the formation of the trion, which was predicted by a recent theoretical study on monolayer phosphorene(*36*) and also well-established in TMD monolayers(*6, 49*) as well as conventional semiconductors(*50, 51*). An exciton can combine with an additional hole (electron) to form a positive (negative) trion. The presence of free carriers enables the optical transition to directly create the trion, while the optical transition for the neutral exciton is suppressed due to a combination of Pauli blocking and dielectric screening. Following photoexcitation, the population of photoexcited excitons quickly relaxes to the trion, which is lower in energy. Such modification of the optical transition and relaxation dynamics of exciton explains our optical spectra in Fig. 3A and 3D.

In addition, the injection of free holes leads to a linewidth broadening and energy blueshift of optical transitions for the exciton 1s and 2s states (Fig. 3D). The scattering with holes decreases the lifetime of excitons which increases the linewidth of the resonances. The blueshift of the resonance energy is due to a combination of Pauli blocking and screening of Coulomb interaction. In the single-particle band picture, the occupation of hole states blocks the interband transitions at the band edge which participate in the formation of the exciton states. This leads to an energy blueshift of the quasi-particle gap and exciton excitation energy. On the other hand, the dielectric screening of Coulomb interaction reduces both the quasiparticle bandgap and the exciton binding energy, which also modifies the exciton resonance energy. Typically, the overall effect results in an energy blueshift for the exciton 1s state, which is consistent with our result(*7, 49*).

The optical spectra for the electron-doped case are investigated by the application of a gate voltage from 0.2 V to 1.5 V with polarization along the armchair direction. The photoluminescence spectra

(Fig. 3B) show somewhat similar behavior to the hole doping case where the exciton peak diminishes accompanied by the emergence of a new peak at ~ 1.09 eV. However, ΔR/R (Fig. 3E) shows a stark difference. We do not observe any noticeable emergence of a new resonance at ~ 1.09 eV corresponding to the luminescence peak. The majority of the 1s state resonance shows only a continuous redshift while leaving a small shoulder at higher energy as shown for $V_B$ = 1.0 V and 1.2 V. At $V_B$ = 1.5 V, as a representative example for the electron-doped case, the magnitudes of the luminescence peak at ~ 1.09 eV and ΔR/R peak at ~ 1.116 eV show the same polarization dependence (red empty circles in Fig. 3C and red filled circles in Fig. 3F, respectively) as the optical resonances for the neutral and hole-doped cases. This gate-dependent behavior cannot be explained by the conventional model for the trion as in the hole-doped case. This can be potentially explained by the electric field effect on few-layer phosphorene. A thicker-layer phosphorene FET with additional gate controls will be effective to address this anomalous issue, which is of future interest but beyond the scope of this work.

The analysis of ΔR/R (Fig. 3D) allows us to estimate the binding energy of the positive trion, which is defined as the energy required to break it into a free hole and the lowest energy neutral exciton. The exciton 1s state has the two possible spin configurations for the electron and the hole which are the optically-bright singlet (S=0) and the optically-dark triplet (S=1) excitons, labelled as $X_0^{singlet}$ (Red solid line) and $X_0^{triplet}$ (Black solid line) respectively in Fig. 4A. The energy levels of two states are separated by twice the electron-hole exchange energy ($E_{ex}$), and the singlet state is higher in energy than the triplet state. In the absence of a magnetic field or spin-orbit coupling, $X_0^{singlet}$ and $X_0^{triplet}$ are bright and dark, respectively, due to the zero quantum number of spin of the photon. Therefore, the exciton 1s peak at 1.129 eV in Fig. 2A and 3D can be attributed to the optical resonance from $X_0^{singlet}$.

While our optical spectroscopy cannot probe the optically dark $X_0^{triplet}$ state, we determine the triplet exciton energy level with theoretical calculations within the *ab initio* GW-BSE approach (see Materials and Methods)(*38-40*). We find that the exciton singlet-triplet energy splitting in bilayer phosphorene is 10 meV in good agreement with previous calculations(*37*). While the encapsulation leads to a renormalization of the QP bandgap and exciton binding energy, it will not affect the exciton singlet-triplet splitting, which arises from the short-range exchange interaction.

Analogous to the case of carbon nanotubes(52, 53), the major dissociation pathway of the positive trion in bilayer phosphorene, labelled as $X_+$ (Red dashed line) in Fig. 4A, is the following process. $X_+ \to X_0^{triplet} + hole$. The energy separation between $X_0^{singlet}$ and $X_+$ monotonically increases as the doping concentration is raised as shown in Fig. 3D. The energy separation between $X_0^{triplet}$ and $X_+$ follows an identical trend, as the exchange interaction defining the exciton singlet-triplet energy splitting is unscreened and should thus be largely unaffected by doping. Fig. 4B summarizes the energy separation as a function of $V_B$. The increase in the energy separation can be explained by the energy required to dissociate $X_+$ into $X_0^{triplet}$ and a hole at different doping concentration(7). At an infinitesimal concentration of doping, the dissociation simply requires $E_B^{X_+}$. At higher concentrations, additional energy is required for the dissociation to overcome the raised chemical potential for the gas of free holes. Fig. 4B shows ~ 20 meV for the energy separation at the nearly zero concentration which corresponds to the binding energy of a positive trion $E_B^{X_+}$.

**Discussion**

In summary, we realize a FET device of high-quality bilayer phosphorene by fabrication of a van der Waals heterostructure with hBN. ΔR/R and photoluminescence spectra show a strong linear-dichroic optical resonances from anisotropic excitons and positive trions. Despite the dielectric screening from hBN, bilayer phosphorene exhibits tightly-bound excitons and trions with a binding energy of roughly 100 meV and 20 meV, respectively. Our work shows exciting possibilities for exploring many-body physics and novel optoelectronic applications based on excitons with two-fold anisotropy.

**Materials and Methods**

**FET device fabrication of hBN-encapsulated bilayer phosphorene**: The van der Waals heterostructure of bilayer phosphorene, hBN and few-layer graphene are prepared with a polyethylene terephthalate (PET) stamp by the dry transfer technique(*31*). All procedures are done in a glove box with a nitrogen environment where oxygen and moisture levels are lower than 1 ppm in order to avoid sample degradation in air. Thickness of phosphorene crystals is first identified by the optical contrast of a microscope image(*21*), which is followed by the detailed spectroscopic characterization. Few-layer graphene and ~ 20 nm-thick hBN flakes are separately exfoliated onto a silicon substrate with 90 nm oxide layer. To fabricate the encapsulated bilayer phosphorene FET, we use the polyethylene terephthalate (PET) stamp to pick up the hBN flakes, few-layer graphene and bilayer phosphorene in sequence with accurate alignment based on an optical microscope. The PET stamp with the heterostructure is then stamped onto sapphire substrate with pre-patterned Ti/Au electrodes. The polymer and samples are heated up at 70 ℃ for the pick-up and 130 ℃ for the stamp process, respectively. Finally, we dissolve the PET in dichloromethane for 12 hours at room temperature.

**Reflection contrast and photoluminescence spectra measurement:** For reflection contrast spectra, a tungsten lamp is used as the broadband white light source. The polarization of incident light was controlled using a broadband half-wave plate and a calcite polarizer. The incident light was focused onto the bilayer phosphorene in a home-built microscopy setup, and the reflected light was collected and analyzed in a spectrometer equipped with the liquid-nitrogen-cooled InGaAs array detector. For photoluminescence spectra, the sample is excited by a unpolarized laser with excitation energy at 1.96 eV. Polarization of outcoming luminescence signal is analyzed by the combination of a broadband half-wave plate and a calcite polarizer in front of the spectrometer. All our optical measurement is carried out at temperature 10 K.

**Calculation of the exciton exchange energy and Rydberg series**: We first performed density functional theory (DFT) calculations in the generalized gradient approximation using the Quantum Espresso code(*54*). We relaxed the geometry of freestanding bilayer phosphorene in a supercell arrangement using a plane wave basis with norm conserving PBE(*55*) pseudopotentials with a van der Waals correction(*56, 57*) and a 55 Ry wave function cutoff. A large vacuum was included between repeated supercells in the aperiodic direction so that 99% of the charge density was contained in half of each supercell.

The GW-Bethe-Salpeter equation (GW-BSE) calculation was done with the BerkeleyGW code(*38*). We performed a one-shot GW calculation to obtain the quasiparticle (QP) bandstructure. The dynamical screening effects were accounted for with the Hybertsen-Louie generalized plasmon pole (HL-GPP) model(*39*). We used a 14×10×1 k-grid to sample the Brillouin zone with additional non-uniform sampling points for an effective sampling of 100x80x1(*58*), included plane-wave components up to a cutoff of 15 Ry in the dielectric matrix, and included unoccupied states with energy up to 10 Ry to converge the QP energies to better than 0.1 eV. The static remainder technique was used to speed up convergence with respect to unoccupied states(*59*). A truncated Coulomb interaction was used to prevent spurious interactions between periodic images(*60*). We solved the BSE(*40*) on a 100×80×1 k-grid, which converges the excitation energies to better than 0.1 eV, and included valence and conduction bands involved in transitions up to 3 eV.

To include the effect of encapsulation, we then numerically solved a modified Wannier-Mott model with an anisotropic kinetic energy (fit to the GW bilayer black phosphorus band structure) and spatially dependent screening following Rytova and Keldysh(*41, 42*), which treats the heterostructure as finite-sized slabs of constant dielectric material. In the limit where the dielectric constant of the material and the surrounding environment are similar, as in the case of black phosphorous and hBN, the Rytova-Keldysh potential reduces to the Coulomb potential

$$V_{RK}(s) = \frac{e}{\epsilon s}$$

, where s is the in-plane radial coordinate and $\epsilon$ is a uniform effective dielectric constant fit to our GW-BSE calculation. Hence, the difference between our model and the isotropic 2D hydrogenic model arises primarily from the anisotropy of the black phosphorus band structure.

**References and Notes**

1. Z. Wang *et al.*, Evidence of high-temperature exciton condensation in two-dimensional atomic double layers. *Nature* **574**, 76-80 (2019).
2. C. Jin *et al.*, Observation of moiré excitons in WSe2/WS2 heterostructure superlattices. *Nature* **567**, 76-80 (2019).
3. K. Tran *et al.*, Evidence for moiré excitons in van der Waals heterostructures. *Nature* **567**, 71-75 (2019).
4. K. L. Seyler *et al.*, Signatures of moiré-trapped valley excitons in MoSe2/WSe2 heterobilayers. *Nature* **567**, 66-70 (2019).
5. K. F. Mak, J. Shan, Photonics and optoelectronics of 2D semiconductor transition metal dichalcogenides. *Nature Photonics* **10**, 216-226 (2016).
6. G. Wang *et al.*, Colloquium: Excitons in atomically thin transition metal dichalcogenides. *Reviews of Modern Physics* **90**, 021001 (2018).
7. K. F. Mak *et al.*, Tightly bound trions in monolayer MoS2. *Nature Materials* **12**, 207-211 (2013).
8. J. S. Ross *et al.*, Electrical control of neutral and charged excitons in a monolayer semiconductor. *Nature Communications* **4**, 1474 (2013).
9. Y. You *et al.*, Observation of biexcitons in monolayer WSe2. *Nature Physics* **11**, 477-481 (2015).
10. K. Hao *et al.*, Neutral and charged inter-valley biexcitons in monolayer MoSe2. *Nature Communications* **8**, 15552 (2017).
11. Z. Li *et al.*, Revealing the biexciton and trion-exciton complexes in BN encapsulated WSe2. *Nature Communications* **9**, 3719 (2018).
12. A. Steinhoff *et al.*, Biexciton fine structure in monolayer transition metal dichalcogenides. *Nature Physics* **14**, 1199-1204 (2018).
13. S.-Y. Chen, T. Goldstein, T. Taniguchi, K. Watanabe, J. Yan, Coulomb-bound four- and five-particle intervalley states in an atomically-thin semiconductor. *Nature Communications* **9**, 3717 (2018).
14. A. Raja *et al.*, Coulomb engineering of the bandgap and excitons in two-dimensional materials. *Nature Communications* **8**, 15251 (2017).
15. M. I. B. Utama *et al.*, A dielectric-defined lateral heterojunction in a monolayer semiconductor. *Nature Electronics* **2**, 60-65 (2019).
16. D. Y. Qiu, F. H. da Jornada, S. G. Louie, Environmental Screening Effects in 2D Materials: Renormalization of the Bandgap, Electronic Structure, and Optical Spectra of Few-Layer Black Phosphorus. *Nano Letters* **17**, 4706-4712 (2017).
17. L. Li *et al.*, Black phosphorus field-effect transistors. *Nature Nanotechnology* **9**, 372-377 (2014).

18. F. Xia, H. Wang, Y. Jia, Rediscovering black phosphorus as an anisotropic layered material for optoelectronics and electronics. *Nature Communications* **5**, 4458 (2014).
19. A. S. Rodin, A. Carvalho, A. H. Castro Neto, Excitons in anisotropic two-dimensional semiconducting crystals. *Physical Review B* **90**, 075429 (2014).
20. A. Chaves, M. Z. Mayers, F. M. Peeters, D. R. Reichman, Theoretical investigation of electron-hole complexes in anisotropic two-dimensional materials. *Physical Review B* **93**, 115314 (2016).
21. L. Li *et al.*, Direct observation of the layer-dependent electronic structure in phosphorene. *Nature Nanotechnology* **12**, 21-25 (2017).
22. G. Zhang *et al.*, Infrared fingerprints of few-layer black phosphorus. *Nature Communications* **8**, 14071 (2017).
23. X. Chen *et al.*, Widely tunable black phosphorus mid-infrared photodetector. *Nature Communications* **8**, 1672 (2017).
24. C. Chen *et al.*, Widely tunable mid-infrared light emission in thin-film black phosphorus. *Science Advances* **6**, eaay6134 (2020).
25. S. Kim *et al.*, Thickness-controlled black phosphorus tunnel field-effect transistor for low-power switches. *Nature Nanotechnology* **15**, 203-206 (2020).
26. X. Wang *et al.*, Highly anisotropic and robust excitons in monolayer black phosphorus. *Nature Nanotechnology* **10**, 517-521 (2015).
27. R. Xu *et al.*, Extraordinarily Bound Quasi-One-Dimensional Trions in Two-Dimensional Phosphorene Atomic Semiconductors. *ACS Nano* **10**, 2046-2053 (2016).
28. G. Zhang *et al.*, Determination of layer-dependent exciton binding energies in few-layer black phosphorus. *Science Advances* **4**, eaap9977 (2018).
29. J. Yang *et al.*, Optical tuning of exciton and trion emissions in monolayer phosphorene. *Light: Science & Applications* **4**, e312-e312 (2015).
30. R. Tian *et al.*, Observation of excitonic series in monolayer and few-layer black phosphorus. *Physical Review B* **101**, 235407 (2020).
31. L. Wang *et al.*, One-Dimensional Electrical Contact to a Two-Dimensional Material. *Science* **342**, 614 (2013).
32. D. Rhodes, S. H. Chae, R. Ribeiro-Palau, J. Hone, Disorder in van der Waals heterostructures of 2D materials. *Nature Materials* **18**, 541-549 (2019).
33. A. Raja *et al.*, Dielectric disorder in two-dimensional materials. *Nature Nanotechnology* **14**, 832-837 (2019).
34. M. Buscema, D. J. Groenendijk, G. A. Steele, H. S. J. van der Zant, A. Castellanos-Gomez, Photovoltaic effect in few-layer black phosphorus PN junctions defined by local electrostatic gating. *Nature Communications* **5**, 4651 (2014).
35. Y. Li, T. F. Heinz, Two-dimensional models for the optical response of thin films. *2D Materials* **5**, 025021 (2018).

36. T. Deilmann, K. S. Thygesen, Unraveling the not-so-large trion binding energy in monolayer black phosphorus. *2D Materials* **5**, 041007 (2018).
37. V. Tran, R. Fei, L. Yang, Quasiparticle energies, excitons, and optical spectra of few-layer black phosphorus. *2D Materials* **2**, 044014 (2015).
38. J. Deslippe *et al.*, BerkeleyGW: A massively parallel computer package for the calculation of the quasiparticle and optical properties of materials and nanostructures. *Computer Physics Communications* **183**, 1269-1289 (2012).
39. M. S. Hybertsen, S. G. Louie, Electron correlation in semiconductors and insulators: Band gaps and quasiparticle energies. *Physical Review B* **34**, 5390-5413 (1986).
40. M. Rohlfing, S. G. Louie, Electron-hole excitations and optical spectra from first principles. *Physical Review B* **62**, 4927-4944 (2000).
41. L. V. Keldysh, Coulomb interaction in thin semiconductor and semimetal films. *Soviet Journal of Experimental and Theoretical Physics Letters* **29**, 658 (1979).
42. N. S. Rytova, The screened potential of a point charge in a thin film. *Moscow university physics bulletin* **3**, 18 (1967).
43. D. Y. Qiu, F. H. da Jornada, S. G. Louie, Optical Spectrum of MoS2: Many-Body Effects and Diversity of Exciton States. *Physical Review Letters* **111**, 216805 (2013).
44. A. Chernikov *et al.*, Exciton Binding Energy and Nonhydrogenic Rydberg Series in Monolayer WS2. *Physical Review Letters* **113**, 076802 (2014).
45. H. M. Hill *et al.*, Observation of Excitonic Rydberg States in Monolayer MoS2 and WS2 by Photoluminescence Excitation Spectroscopy. *Nano Letters* **15**, 2992-2997 (2015).
46. D. Y. Qiu, F. H. da Jornada, S. G. Louie, Screening and many-body effects in two-dimensional crystals: Monolayer MoS2. *Physical Review B* **93**, 235435 (2016).
47. P. J. Gielisse *et al.*, Lattice Infrared Spectra of Boron Nitride and Boron Monophosphide. *Physical Review* **155**, 1039-1046 (1967).
48. O. Madelung, Semiconductors - Basic Data. *Springer: Berlin*, (1996).
49. A. Chernikov *et al.*, Electrical Tuning of Exciton Binding Energies in Monolayer WS2. *Physical Review Letters* **115**, 126802 (2015).
50. S. W. K. N. Peyghambarian, and A. Mysyrowicz, Introduction to semiconductor optics. *Prentice-Hall, Englewood Cliffs, NJ*, (2009).
51. H. H. a. S. W. Koch, Quantum theory of the optical and electronic properties of semiconductors 5th ed. *World Scientific, Singapore*, (2009).
52. S. M. Santos *et al.*, All-Optical Trion Generation in Single-Walled Carbon Nanotubes. *Physical Review Letters* **107**, 187401 (2011).
53. J. S. Park *et al.*, Observation of Negative and Positive Trions in the Electrochemically Carrier-Doped Single-Walled Carbon Nanotubes. *Journal of the American Chemical Society* **134**, 14461-14466 (2012).

54. P. Giannozzi *et al.*, QUANTUM ESPRESSO: a modular and open-source software project for quantum simulations of materials. *Journal of Physics: Condensed Matter* **21**, 395502 (2009).
55. J. P. Perdew, K. Burke, M. Ernzerhof, Generalized Gradient Approximation Made Simple. *Physical Review Letters* **77**, 3865-3868 (1996).
56. S. Grimme, Semiempirical GGA-type density functional constructed with a long-range dispersion correction. *Journal of Computational Chemistry* **27**, 1787-1799 (2006).
57. V. Barone *et al.*, Role and effective treatment of dispersive forces in materials: Polyethylene and graphite crystals as test cases. *Journal of Computational Chemistry* **30**, 934-939 (2009).
58. F. H. da Jornada, D. Y. Qiu, S. G. Louie, Nonuniform sampling schemes of the Brillouin zone for many-electron perturbation-theory calculations in reduced dimensionality. *Physical Review B* **95**, 035109 (2017).
59. J. Deslippe, G. Samsonidze, M. Jain, M. L. Cohen, S. G. Louie, Coulomb-hole summations and energies for $GW$ calculations with limited number of empty orbitals: A modified static remainder approach. *Physical Review B* **87**, 165124 (2013).
60. S. Ismail-Beigi, Truncation of periodic image interactions for confined systems. *Physical Review B* **73**, 233103 (2006).

**Acknowledgement**:

**Funding:** J.K., S.Y., T.K. S.-B.S. acknowledge the support from the National Research Foundation of Korea grant (No. 2017R1C1B2012729 and NRF-2020R1A4A1018935) and POSCO Steel Science Program. B.J.K. acknowledges the support from the Institute for Basic Science (IBS-R014-A2). S.-Y. S. and M.-H.J acknowledges the support the Institute for Basic Science (IBS), under Project Code IBS-R014-A1. S.-H.S. and G.-H.L. acknowledge the support from the National Research Foundation of Korea (NRF-2020R1C1C1013241). K.W. and T.T. acknowledge support from the Elemental Strategy Initiative conducted by the MEXT, Japan, Grant Number JPMXP0112101001, JSPS KAKENHI Grant Number JP20H00354 and the CREST(JPMJCR15F3), JST. D.Y.Q. acknowledges resources from a user project at the Molecular Foundry supported by the Office of Science, Office of Basic Energy Sciences, of the U.S. Department of Energy under Contract No. DE-AC02-05CH11231. Computational resources were provided by resources of the National Energy Research Scientific Computing Center (NERSC), a DOE Office of Science User Facility supported by the Office of Science of the U.S. Department of Energy under Contract No. DE-AC02-05CH11231, which provided computational resources for the calculation on benzene.

**Author contributions**: J.K. conceived the project. S.Y., T.K., S.-Y.S., S.-H.S., S.-B.S., G.-H.L and M.-H.J. fabricated the FET device of hBN-encapsulated bilayer phosphorene. S.Y., T.K., B.J.K., D.Y.Q. and J.K. obtained and analysed optical spectra. D.Y.Q. performed the ab-initio calculation on the exciton structure. K. W. and T. T. grew hBN crystals. All authors discussed and wrote the manuscript together.

# Figures

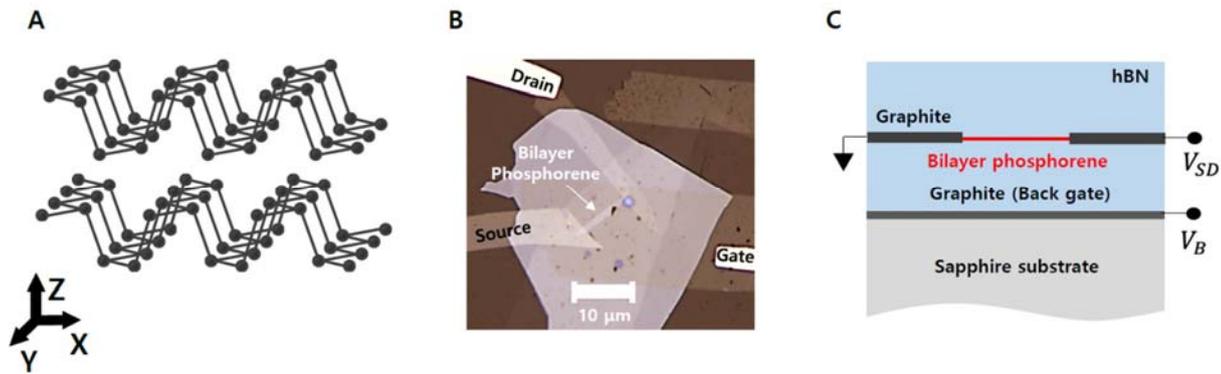

**Fig. 1 | hBN-encapsulated bilayer phosphorene in FET geometry. (A)** Crystal structure of bilayer phosphorene. Phosphorus atoms form a puckered honeycomb lattice with two distinct crystalline axes along the X (armchair) and Y (zig-zag) directions. **(B)** Optical microscope image of the representative bilayer phosphorene device. Bilayer phosphorene (white arrow) is encapsulated by hBN. The scale bar corresponds to 10 μm. **(C)** The schematic of the FET device in **(B)**. In addition to hBN encapsulation, two flakes of few-layer graphene are in contact to bilayer phosphorene for the source and drain electrodes. The additional few-layer graphene at the bottom is used as the gate electrode. For the optical measurement, both the source and drain electrodes are grounded, and the bottom gate voltage ($V_B$) is varied from -1.5V to +1.5V. For the transport measurement, 300 mV is applied for $V_{SD}$ at the source electrode.

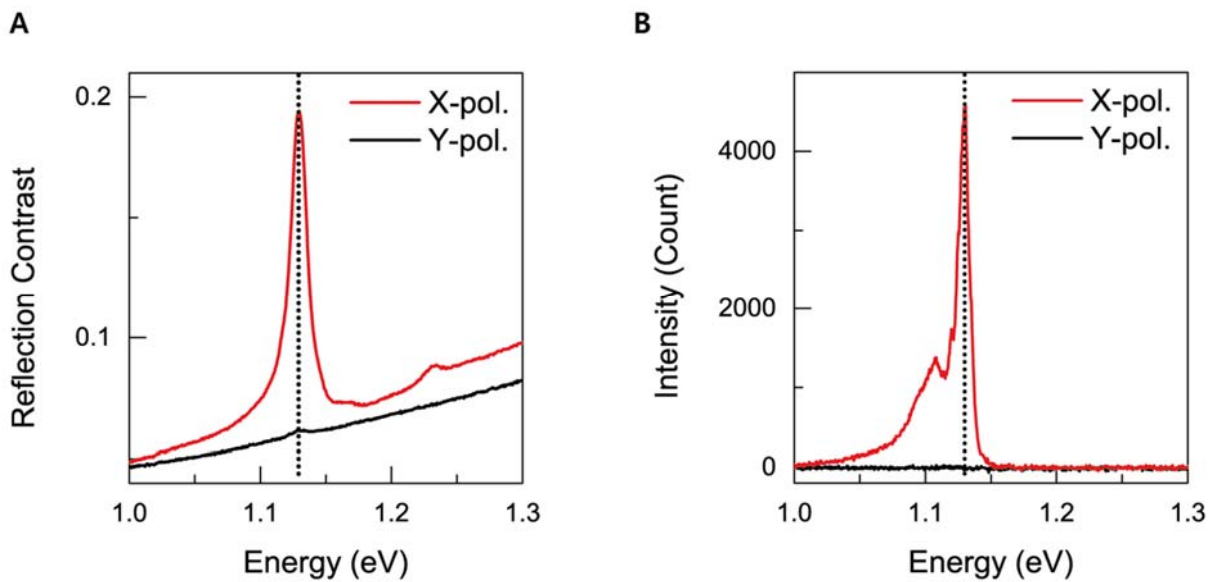

**Fig. 2 | Optical spectra of the anisotropic exciton for the undoped system. (A)** The polarization-resolved ΔR/R. For the polarization along the arm-chair direction (Red solid line labelled as X-pol.), the prominent resonances at 1.129 eV and 1.229 eV originate from the exciton 1s and 2s state. For the polarization along the zig-zag direction (Black solid line labelled as Y-pol.), ΔR/R shows an overall broad background from the imaginary part of the optical conductivity in bilayer phosphorene. **(B)** The polarization-resolved photoluminescence spectra with the unpolarized laser excitation at 1.96 eV. The luminescence shows strong signal along X-polarization at 1.129 eV which is exactly the same energy of the resonance in **(A).** as marked with the black dashed line. The luminescence does not show any signal along Y-polarization.

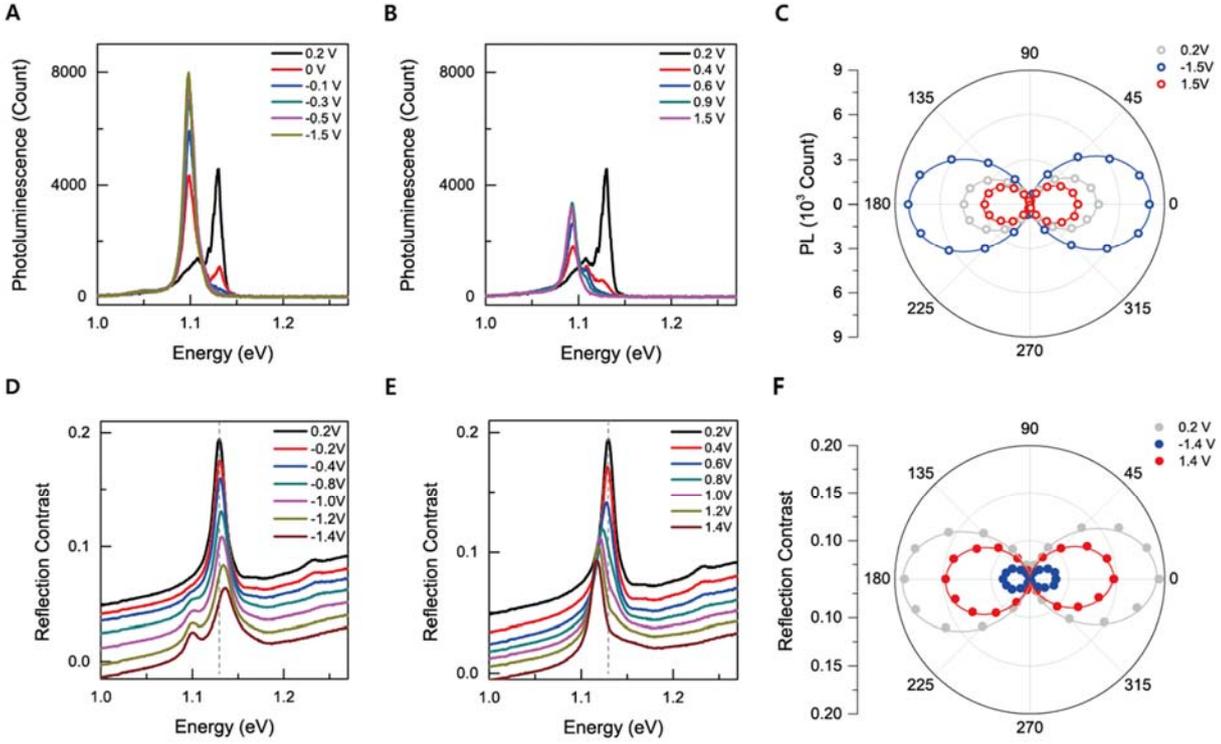

**Fig. 3 | Optical spectra of the anisotropic excitons with gate control. (A)** and **(B)** Photoluminescence spectra with polarization along the armchair direction for the hole-doped **(A)** and electron-doped **(B)** cases. **(C)** Intensity of photoluminescence at peak energy as a function of polarization angle $\theta$. Grey-, blue-, and red-empty circles show data at $V_B$ = 0.2 V (the charge-neutral case), -1.5 V (the hole-doped case), and 1.5 V (the electron-doped case), respectively. **(D)** and **(E)** ΔR/R with polarization along the armchair direction for the hole-doped **(D)** and electron-doped **(E)** cases. **(F)** ΔR/R at peak energy as a function of polarization angle $\theta$. Grey-, blue-, and red-filled circles show data for peaks at 1.129 eV ($V_B$ = 0.2 V), 1.1 eV ($V_B$ = -1.4 V), and 1.116 eV ($V_B$ = 1.4 V), respectively. For the hole-doped case, the exciton 1s peak at 1.129 eV diminishes, and a new peak emerges at ~ 1.1 eV as the gate voltage is varied from 0.2 V to -1.5 V in both photoluminescence **(A)** and ΔR/R spectra **(D)**. For the electron-doped case, on the other hand, the photoluminescence spectra **(B)** show a new peak at ~ 1.09 eV, but ΔR/R spectra **(E)** show continuous redshift of the exciton 1s state while leaving a small shoulder at higher energy as shown for $V_B$ = 1.0 V and 1.2 V. All of the optical resonances show a nearly perfect $cos^2\theta$ pattern (Grey, blue and red solid lines) with the minimum at $\theta = 90°$ (Y-pol.), consistent with the fact that dipole interaction is forbidden for the polarization along the zig-zag direction by symmetry at the band edge.

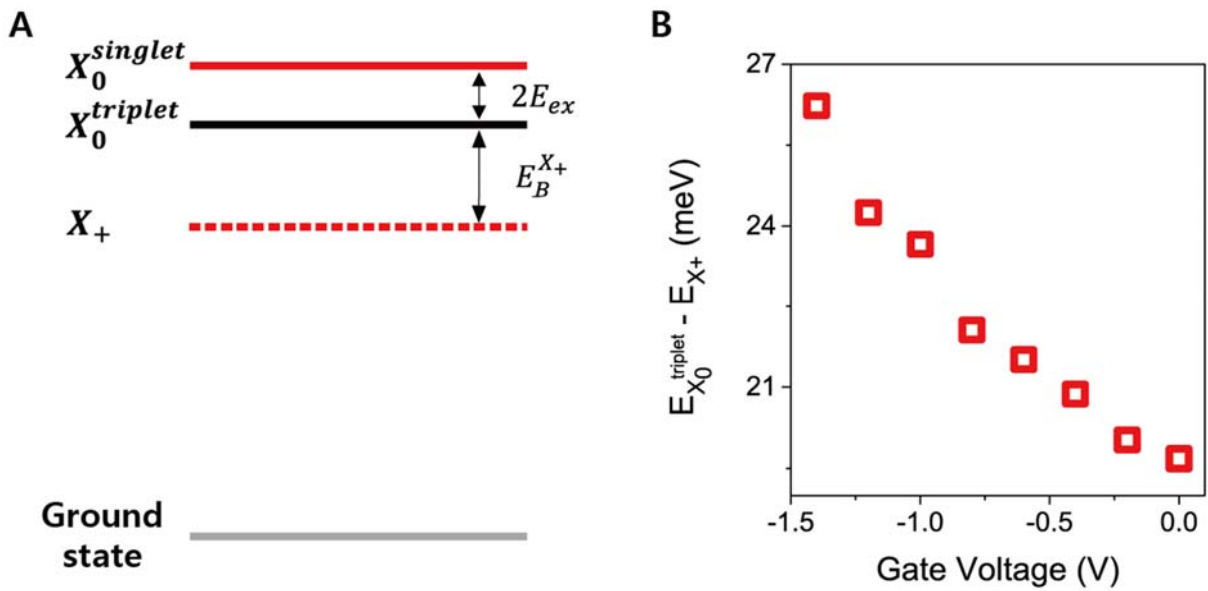

**Fig. 4 | The gate-control of exciton energy structure. (A)** The schematic of the energy levels for excitons and trions. The spin-singlet and triplet of the exciton 1s states are labelled as $X_0^{singlet}$ (Red solid line) and $X_0^{triplet}$ (Black solid line) and are separated by twice the electron-hole exchange energy ($E_{ex}$). The positive trion is labelled as $X_+$ (Red dashed line) is located below $X_0^{triplet}$ by the binding energy ($E_B^{X_+}$). **(B)** The energy separation of optical resonances between $X_0^{triplet}$ and $X_+$ with the function of $V_B$. The energy separation monotonically increases with the doping concentration since the dissociation requires the additional energy for the raised chemical potential for the gas of free holes.